\documentclass[journal]{IEEEtranTIE}
\usepackage{graphicx}

\usepackage{picinpar}
\usepackage{amsmath}
\usepackage{url}
\usepackage{flushend}
\usepackage[latin1]{inputenc}
\usepackage{colortbl}
\usepackage{soul}
\usepackage{multirow}
\usepackage{pifont}
\usepackage{color}
\usepackage{alltt}
\usepackage[hidelinks]{hyperref}
\usepackage{enumerate}
\usepackage{siunitx}
\usepackage{breakurl}
\usepackage{epstopdf}
\usepackage{pbox}
\usepackage{amsmath}
\usepackage{cite}
\usepackage{amsmath,amssymb,amsfonts,amsthm}
\usepackage{hyperref}
\usepackage[table]{xcolor}
\usepackage{textcomp}
\usepackage{placeins}
\usepackage{placeins}
\usepackage{amsthm}
\usepackage{comment}

\graphicspath{{VISIO0511/}{../pdf/}{../jpeg/}{./Photo/}}
\DeclareGraphicsExtensions{.pdf,.jpeg,.png}

\begin{document}
	
\title{Symmetric Sliding-Mode  Control of Grid-Forming Inverters With Precision Region Under AC and DC Sides Varying}
\author{
\vskip 1em
   Qianxi~Tang, \emph{Graduate Student Member,~IEEE},
   Li~Peng, \emph{Senior Member,~IEEE},
    Xuefeng~Wang,
    Xinchen~Yao
	\vskip 1.8em
	\large{\textbf{Post Conference Paper}}
	\small{\textbf{\color{red} [DELETE THIS LINE FROM YOUR ACCEPTED FINAL SUBMISSION]} }

\thanks{
	This work has been submitted to the IEEE for possible publication. Copyright may be transferred without notice, after which this version may no longer be accessible. 
	 (Corresponding author: Li~Peng.)
 
Qianxi~Tang, Li~Peng, Xuefeng~Wang, and  Xinchen~Yao
are with the State Key Laboratory of Advanced Electromagnetic Engineering and Technology, School of Electrical and Electronic Engineering, Huazhong University of Science and Technology, Wuhan 430074, China (e-mail: qianxi@hust.edu.cn; pe105@mail.hust.edu.cn;
xuefeng090@hust.edu.cn; yaoxc@hust.edu.cn).
}

}

\maketitle
	
\begin{abstract}
	Voltage regulation under conventional grid-forming controllers is tightly coupled to power sharing and dc-link dynamics. Consequently, its tracking accuracy deteriorates during grid faults, sudden power sharing changes, or dc-bus voltage varying. To address this issue, a symmetric sliding-mode control (SSMC) method is developed and its voltage precision region is derived. It illustrates how much ac-side power dynamics and dc-link voltage varying can be decoupled from the voltage regulation task, which helps predict when an abnormal entangling appears. While conventional sliding-mode controls address voltage-tracking error through complex sliding surface designs, repetitive correction techniques or special reaching laws, this work identifies that the error at power-line frequency primarily stem from the asymmetry property of inverters with the delay effect and the computational inaccuracy. Guided by this insight, an asymmetry compensation structure is proposed, which avoids added design complexity and directly mitigates voltage tracking error. Furthermore, the control design is supported by a physical and quantitative explanation, aiding in parameter tuning. Simulation and experimental results demonstrate that the proposed method achieves faster tracking responses while maintaining robust and more accurate tracking under both dc-link voltage and ac-side current variations. Conventional grid-forming and classical sliding-mode controllers, which handle these variations separately, cannot match this combined speed and robustness. Furthermore, the voltage precision region is explicitly verified.

\end{abstract}


\markboth{IEEE TRANSACTIONS ON INDUSTRIAL ELECTRONICS}%
{}

\definecolor{limegreen}{rgb}{0.2, 0.8, 0.2}
\definecolor{forestgreen}{rgb}{0.13, 0.55, 0.13}
\definecolor{greenhtml}{rgb}{0.0, 0.5, 0.0}

\section{Introduction}

\IEEEPARstart{G}{rid}-forming inverters are interfaces between the modern grid and renewable or power storage devices to regulate the ac-side voltage. Because of the modern grid which is composed of more distributed power sources, fewer large-inertia synchronous generators and intermittent nature of wind and solar resources etc., dc-link voltage and ac-side power dynamics become more sophisticated. Therefore, how this grid-forming interface technology stably provides voltage regulation service, regardless of  the significantly fluctuating from dc-link voltage and ac-side current, becomes a serious challenge.
Therefore, some new voltage controllers are proposed to achieve  good voltage tracking under  significantly varying
operating-point condition and wide stability-range robustness \cite{1,2,3}. 
Nevertheless, voltage tracking is slow and exists error in low-frequency and voltage-forming accuracy is coupled with power injection to grid, which might damage the performance of synchronous loop. Furthermore, these methods haven't provided the voltage precision region related to its dc-voltage level and ac-current changing rate explicitly. This omission makes it difficult to assess under what conditions a voltage regulation strategy will fail and what the voltage trajectory could be when the inverter is outside the voltage precision region.  As a result, the stability of a GFMI becomes an issue that the voltage loop and the synchronization loop are intricately interacted\cite{4}, which is hard to be analyzed. It remains unclear whether failure arises primarily from poor voltage tracking or loss of synchronization, and in many cases, failures are attributed to the both simultaneously without identifying the dominant cause \cite{5}. The confusion is further compounded in studies that assume ideal voltage control, i.e., perfect voltage tracking \cite{6}. 

While some recent methods attempt to regulate the dc-link voltage to support ac-side voltage stability under large states varying \cite{7}, this often comes at the cost of reduced voltage regulation capability at the point of common coupling (PCC). For instance, strict dc-voltage level regulation can limit the controllability of either the voltage magnitude or the frequency \cite{8,9}. This trade-off raises a critical question: rather than enforcing tight dc-voltage regulation, why not consider how much dc-voltage variation the voltage controller can be tolerated by GFMI while GFMI is still maintaining robust and accurate ac-voltage regulation? 

To achieve accurate and robust voltage tracking, and to analyze the precision boundary condition, a sliding-mode control (SMC) strategy rooted in nonlinear system analysis is employed. Because the scheme derives the precision region in closed form, the tracking accuracy and convergence speed can be rigorously quantified. Furthermore, the analysis shows that the voltage precision region is a time function of dc-voltage level and ac-current changing rate, which means the answer of how much dc-voltage variation the voltage controller can tolerate while still maintaining robust and accurate ac-voltage regulation is given in this condition. Moreover, conventional sliding-mode voltage control for grid-forming inverters rarely connect the discrete switching action of the power stage with the reaching law.  To address it, designers introduce higher-order or adaptive sliding surfaces and bespoke reaching laws, but the resulting schemes still fall short of microsecond-level settling and leave a discernible 50/60 Hz residual \cite{10752352,8391293,8240975}.  The proliferation of auxiliary gains also obscures the physical link between each parameter and the inverter's energy-conversion process, making control-parameter choosing hard.

Motivated by the above problems, this paper proposes a symmetric sliding-mode voltage control scheme. It also offers a nonlinear analysis based control-parameter design. The main contributions are as follows:
\begin{enumerate}[1)]
	\item The explicit precision region is introduced. In this region, the tracking performance of SSMC, which is faster and more precise, lowers the coupling or entangling between synchronous loop and voltage inner loop. So, GFMI robustness and  stability can be enhanced during large operating-region changing.
	\item The proposed SSMC improves voltage tracking under simultaneous dc-link-voltage and ac-side-current variations. Because of its physically grounded analysis of asymmetry, its control complexity is lower than some traditional advanced sliding-mode controls. 
	\item 
    The controller's nonlinear analysis based parameter choosing guarantees accurate voltage formation even under large ac and dc variations. It offers straightforward implementation and a transparent, theoretically sound method for parameter tuning.
 \end{enumerate}

\section{Symmetric Sliding-Mode Control With Precision Region}

\begin{figure}[!t]\centering
	\includegraphics[width=8.5cm]{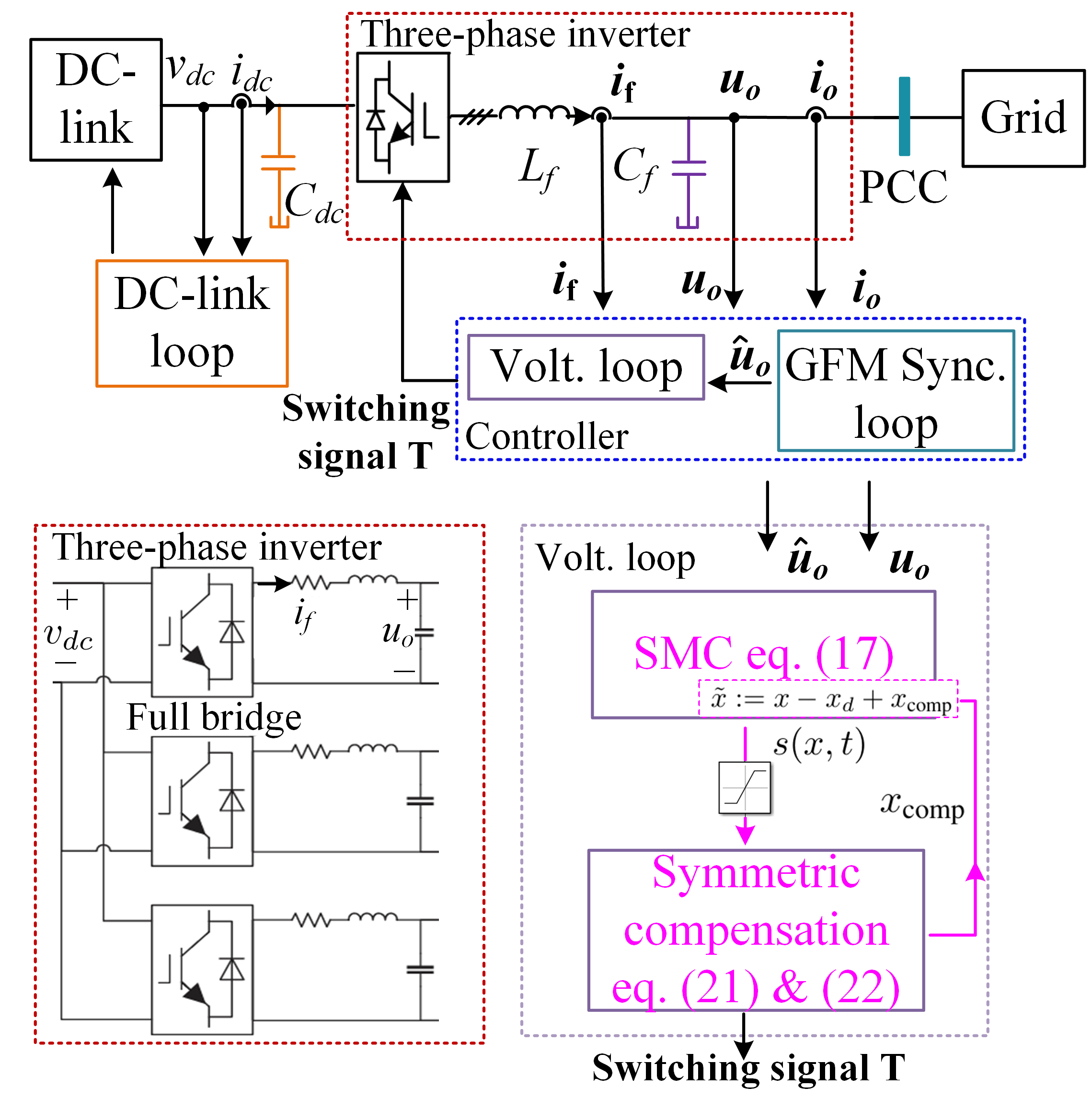}
	\caption{The diagram and control architecture of the GFMI system.}
	\label{FIG_1}
\end{figure}

\begin{figure}[!htbp]
	\includegraphics[width=3in]{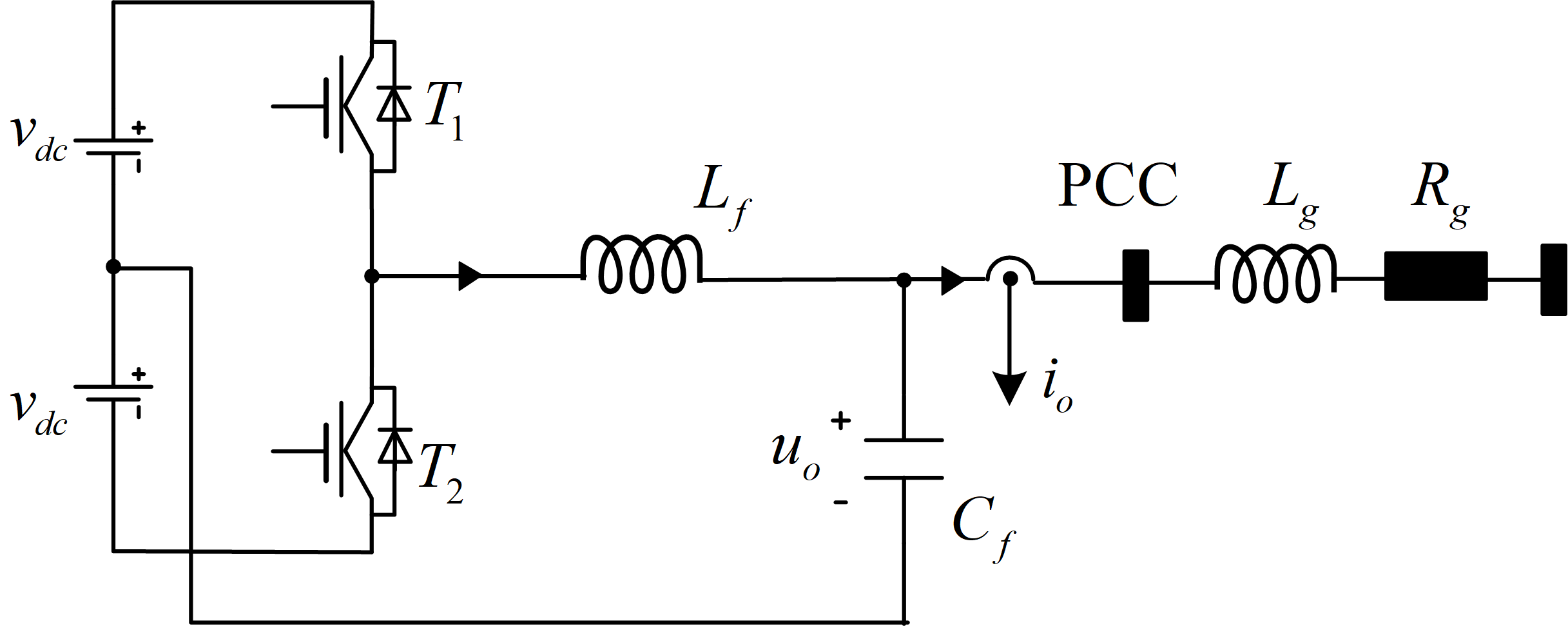}
	\centering
	\caption{The equivalent one phase inverter.}
	\label{FIG_2}
\end{figure}

The diagram and control architecture of the GFMI system is shown in Fig.~\ref{FIG_1}. The topology of the three-phase voltage source inverter is chosen to be the prototype. With three-phase balanced condition, the equivalent single-phase model in Fig.~\ref{FIG_2} is adopted for the analysis \cite{11}. Then, the other two can use the same analysis following below \cite{10}.
Background on sliding-mode control can be found in \cite{Slotine1991AppliedNC,10429007}.


\begin{equation}
		\label{eqd2}
	\begin{aligned}
		&L_f \frac{d i_f}{d t}=T v_{dc}-u_o\\
		&C_f \frac{d u_o}{d t}=i_f-i_o 
	\end{aligned}
\end{equation}

Fig.~\ref{FIG_2} illustrates an inverter system with a dc-link, an LC filter and a transmission line. In its dynamic equation \eqref{eqd2}, the variables are capacitor voltage $u_o$, output current $i_o$, switching variable $T$, dc-link $v_{dc}$. When switch $T_1$ is on, $T=1$. When switch $T_2$ is on, $T=-1$. The variable $i_f$ is now eliminated, the term $Tv_{dc}$ is replaced with control input $u$, and the terms $u_o$ and $i_o$ are combined into one system function $f\left(u_o, i_o\right)$ in \eqref{eqd3}.
\begin{equation}
	\begin{aligned}
		\label{eqd3}
		&\frac{d^2 u_o}{d^2 t}=-\frac{u_o}{L_f C_f}-\frac{1}{C_f} \frac{d i_o}{d t}+\frac{T v_{dc}}{L_f C_f} \\
		&\frac{d^2 u_o}{d^2 t}=-\frac{u_o}{L_f C_f}-\frac{1}{C_f} \frac{d i_o}{d t}+u=f\left(u_o, i_o\right)+u \\
			\end{aligned}
\end{equation}

Note that it's a second order system. The $f$ and $u$ are like forces changing the displacement vector, $u_o$. The $i_o$ here is the current injected into grid side and is known and measured by sensor. 

Now, the sliding-mode control can be derived. First, define the tracking error in \eqref{eqd4}, where $x:=u_o$ and $x_d$, $u_d$ are the desired output-voltage trajectory. Furthermore, define $s$ in \eqref{eqd4}.
\begin{equation}
	\begin{aligned}
		\label{eqd4}
		&\tilde{x}:=x-x_d=u_o-u_d\\
		&s(x, t):=\left(\frac{d}{d t}+\lambda\right) \tilde{x}
		\end{aligned}
\end{equation}

A conclusion \cite{Slotine1991AppliedNC} is that bounds on $s$ can directly translated into bounds on the tracking error vector $\tilde{x}$, when the initial tracking error is 0, given as
\begin{equation}
	\label{eq5}
	\forall t \geq 0,|s(t)| \leq \Phi \Rightarrow \forall t \geq 0,|\tilde{x}| \leq \frac{\Phi}{\lambda}
	\end{equation}
	
	\begin{figure*}[!t]\centering
		\includegraphics[width=0.9\textwidth]{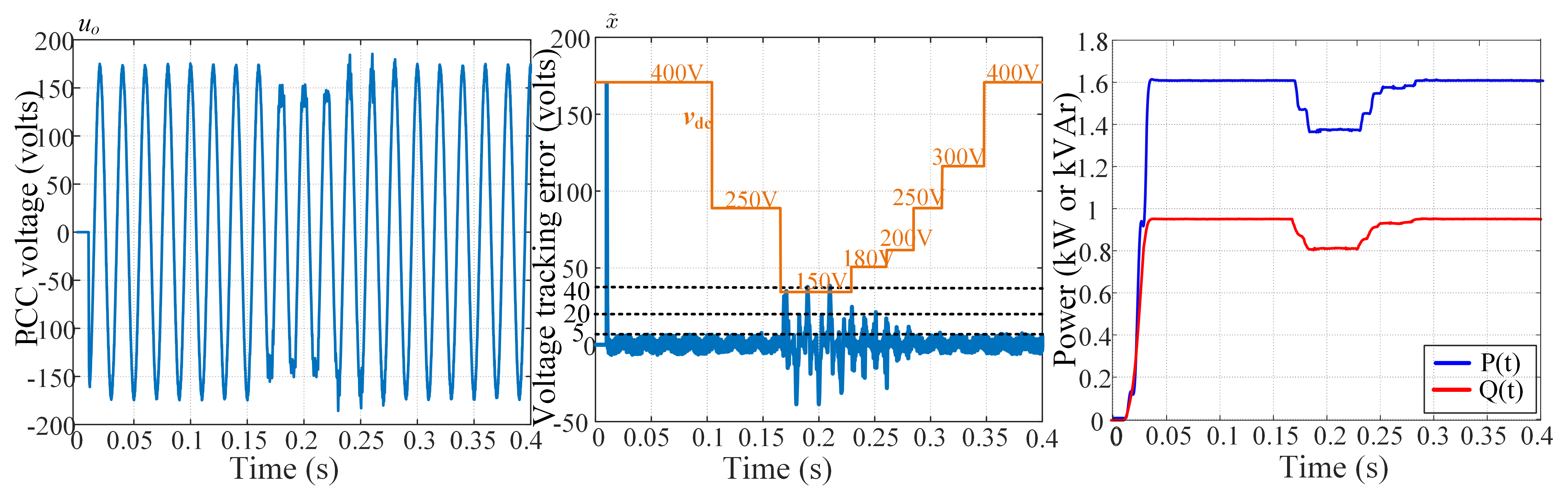}
		\caption{Simulation to examine voltage precision region varying $v_{dc}$  from 150V to 400V. }
		\label{controregion}
	\end{figure*}
	
where $\lambda$ is a strictly positive constant parameter. The physical meaning is that filtering the  $s$ chattering using $\lambda$ makes the $\tilde{x}$ zero. So, the $\lambda$ should be high enough to filter the direct component of  $s$ and low enough to filter out the switching ripple of  $s$, typically  $\lambda=\frac{Switching~frequency}{5}$ \cite{Slotine1991AppliedNC}.
And the initial zero tracking condition\cite{Slotine1991AppliedNC} is guaranteed by 
\begin{equation}
		\label{eq6}
	\frac{1}{2} \frac{d}{d t} s^2 \leq-\eta|s| 
\end{equation}
where the $\eta$ is a strictly positive constant representing the desired lowest converging speed, which guarantees the finite-time convergence. The derivative of $s$ is 
\begin{equation}
	\label{eq61}
	\dot{s}=f-\ddot{x}_d+\lambda \dot{\tilde{x}}+u.
\end{equation}

Using the average model \cite{10329961},  the modulation waveform could be chosen as ideal control input $u_{ideal}$ in \eqref{eq7}.
\begin{equation}
	\label{eq7}
	u_{ideal}=-\hat{f}+\ddot{x}_d-\lambda \dot{\tilde{x}}-(F+\eta) \operatorname{sgn}(s)
\end{equation}
where $|\hat{f}-f| \leq F $. $\hat{f}$ is the estimated model, whereas $f$ is the true, perfectly accurate model. $F$ represents modeling uncertainty \cite{Slotine1991AppliedNC}.  The reason of \eqref{eq7} as control law works, is that the control input is an averaged discontinuous value, which is a function of $v_{dc}$, so the nature of this inverter system is a variable structure system corresponding to the discontinuous switching behaviors of the bridge arms. When $x$ is around a desired trajectory, the forces to change inverter states are discontinuously up and down crossing the desired trajectory to make the system have the freedom to move along the desired trajectory. With this understanding, the  voltage precision region is naturally obtained in \eqref{eq8}, where $g(x)=\frac{x}{L_f C_f}$, to satisfy contraction to $s$ surface condition\cite{464511}.  
\begin{equation}
	\begin{aligned}
		\label{eq8}
		& u=g\left(-v_{dc}\right) \leq-\hat{f}+\ddot{x}_d-\lambda \dot{\tilde{x}}-(F+\eta),s>0 \\
		& u=g\left(+v_{dc}\right) \geq-\hat{f}+\ddot{x}_d-\lambda \dot{\tilde{x}}+(F+\eta), s<0 
		\end {aligned}
	\end{equation}
	
	\begin{proof}
		
		\textit{Case $s>0$.}  
		Apply the first line of \eqref{eq8}:
		\[
		u\;\le\;-\hat f+\ddot x_d-\lambda\dot{\tilde x}-(F+\eta).
		\]
		Substituting into \eqref{eq61}, gives
		\[
		\dot s
		\le (f-\hat f)-(F+\eta)
		\le -\eta,
		\]
		because $|\hat{f}-f| \leq F $.
		
		\textit{Case $s<0$.}  
		Use the second line of \eqref{eq8}:
		\[
		u\;\ge\;-\hat f+\ddot x_d-\lambda\dot{\tilde x}+(F+\eta),
		\]
		which yields
		\[
		\dot s
		\ge (f-\hat f)+(F+\eta)
		\ge +\eta,
		\]
		since $|\hat{f}-f| \leq F $.
		
		Define the Lyapunov candidate \(V=\tfrac12 s^{2}\).  
		Then \(\dot V = s\,\dot s\), and the two cases above imply
		\[
		\dot V\;=\;\frac12\frac{d}{dt}s^{2}\;\le\;-\eta|s|,
		\]
		which is exactly \eqref{eq6}.  
	
		In both cases we have $\dot V=s\,\dot s\le -\eta|s|$. Because $\dot V$ is negative definite outside $s=0$, standard finite-time convergence results for first-order sliding
		surfaces imply reachability in time less than $ |s(0)|/\eta$.  
	\end{proof}
	
Therefore, only when the \eqref{eq8} is satisfied, both the desired converging speed $\eta$ and voltage precision is guaranteed. This boundary condition tells when the voltage regulation control would lose precision. It shows that the voltage precision is related to the  desired trajectory, the output current changing rate, the inductor, modeling uncertainty, required convergence speed and the dc-link voltage level. It also illustrates that the voltage precision depends on the inverter's hardware and the desired performance requirements. It is proved that \eqref{eqd4} is a bijection \cite{12}. So, the controllability of $s$ is the controllability of  $\tilde x$.

\subsection{Voltage Precision Region}


The inverter's ability to follow a voltage reference is restricted by  
the precision window \eqref{eq8}, which can be written as \eqref{eq:VGCRlambda}.

\begin{equation}\label{eq:VGCRlambda}
	\begin{aligned}
		\Bigl|\ddot u_d
		+\frac{u_o}{L_f C_f}
		+\frac{1}{C_f}\frac{d i_o}{d t}
		+\lambda\dot{\tilde x}\Bigr|
		\;\le\;
		\frac{v_{dc}}{L_f C_f}-(F+\eta)
	\end{aligned}
\end{equation}

However, this precision region is not directly related to inverter states, which is not convenient to determine whether the voltage is precise  or not under different application scenarios. If  the  $\dot{\tilde x}$ term is replaced by the inverter states, that goal can be realized.

\paragraph{Step\,1: express $\dot{\tilde x}$ through $s$ and $\tilde x$}
By definition of the sliding surface,
\begin{equation}\label{eq:def_surface}
	s \;=\; \dot{\tilde x}+\lambda\tilde x
	\;\;\Longrightarrow\;\;
	\dot{\tilde x} \;=\; s-\lambda\tilde x .
\end{equation}

\paragraph{Step\,2: apply the triangle inequality}
\begin{equation}\label{eq:tri}
	\bigl|\dot{\tilde x}\bigr|
	\;\le\;
	|s| \;+\; \lambda\,|\tilde x| .
\end{equation}

\paragraph{Step\,3: insert a state bound obtained from the reaching law}
Using the reaching inequality
$\tfrac12\dot s^{2}\le-\eta|s|$
together with~\eqref{eq:def_surface} one show
\begin{equation}\label{eq:x_bound}
	|\tilde x|
	\;\le\;
	\frac{|s|}{\lambda} \;+\; \frac{\eta}{\lambda^{2}}.
\end{equation}

Substituting~\eqref{eq:x_bound} into~\eqref{eq:tri} yields

\begin{equation}
	\label{eq:B}
	\begin{aligned}
		|\dot{\tilde x}|
		&\le
		|s|+\lambda\!\left(\frac{|s|}{\lambda}+\frac{\eta}{\lambda^{2}}\right)\\[2pt]
		&=
		2|s|+\frac{\eta}{\lambda}.
	\end{aligned}
\end{equation}

\paragraph{Step\,4: eliminate $|s(t)|$ with its worst-case decay}
From $\tfrac12\dot s^{2}\le-\eta|s|$ we have
\begin{equation}\label{eq:s_linear}
	|s(t)|\;\le\;|s(t_{0})|-\eta\,(t-t_{0}),
	\qquad
	t_{0}\le t\le t_{0}+|s(t_{0})|/\eta .
\end{equation}


\paragraph{Step\,5: Final form (no $\dot{\tilde x}$ term)}

Inserting~\eqref{eq:s_linear} into~\eqref{eq:B} gives the explicit
time-dependent bound of $|\dot{\tilde x}(t)|$.
\begin{equation}\label{eq:xdot_bound}
	\begin{aligned}
		|\dot{\tilde x}(t)|
		\;&\le\;
		2 \!\bigl(|s(t_{0})|-\eta(t-t_{0})\bigr)
		+\dfrac{\eta}{\lambda},
		\\[-2pt]
		&t_{0}\le t\le t_{0}+\dfrac{|s(t_{0})|}{\eta}.
	\end{aligned}
\end{equation}

Inserting~\eqref{eq:xdot_bound} into~\eqref{eq:VGCRlambda} gives the final form
 of voltage precision region.

\begin{equation}\label{eq:VGCR_final}
	\begin{aligned}
		\left|
		\ddot u_{d}
		+\frac{u_{o}}{L_{f}C_{f}}
		+\frac{1}{C_{f}}\frac{d i_{o}}{d t}
		\right|
		&\;\le\;
		\frac{v_{dc}}{L_{f}C_{f}}
		-(F+2\eta)                                     \\[3pt]
		&\!\!\!
		-\,2\lambda\bigl(|s(t_{0})|-\eta(t-t_{0})\bigr),
	\end{aligned}
\end{equation}
\[
t_{0}\le t\le t_{0}+\frac{|s(t_{0})|}{\eta}.
\]

The transient term $-2\lambda\bigl(|s(t_{0})|-\eta(t-t_{0})\bigr)$ is most negative at $t=t_{0}$ and relaxes \emph{linearly} to zero as $|s(t)|$ collapses; hence a design with a
modest~$\eta$ may momentarily violate the bound at start-up, yet satisfy it
automatically a few switching cycles later.  Conversely, selecting a
larger~$\eta$ widens the static deduction $F+2\eta$ but shortens the
reaching interval ${|s(t_{0})|}/{\eta}$.  Once the surface is around
($s\!\to\!0$) the transient offset disappears and the right hand side of the inequality \eqref{eq:VGCR_final}  reduces to
$\frac{v_{dc}}{L_{f}C_{f}}-(F+2\eta)$. The simulation is given in this section B to check its effectiveness after the control law is provided.   
     

\subsection{The Asymmetry and the Symmetric Control}

Different from the traditional sliding-mode control methods which improve the tracking performance by using advanced sliding surfaces and reaching laws, this paper will find the root of the asymmetry problem causing an error when it is tracking a trajectory. The dynamic of $s$ is in \eqref{eq9} and \eqref{eq10}, when the hysteresis function is used to generate switching signals. The hysteresis function meets \eqref{eq8} outside its band, which guarantees the convergence to the sliding surface.
\begin{equation}
	\label{eq9}
	\dot{s}  =f-\ddot{x}_d+\lambda \dot{\tilde{x}}+\frac{v_{dc} \operatorname{hysterisis}(\frac{-s}{h})}{L_f C_f}
	\end{equation}			
\begin{equation}
	\label{eq10}
		s_n-s_{n-1}=\dot{s}_{n-1} \times t_{\text {di }}
		\end{equation}
\begin{figure*}[!t]\centering
		\includegraphics[width=0.88\textwidth]{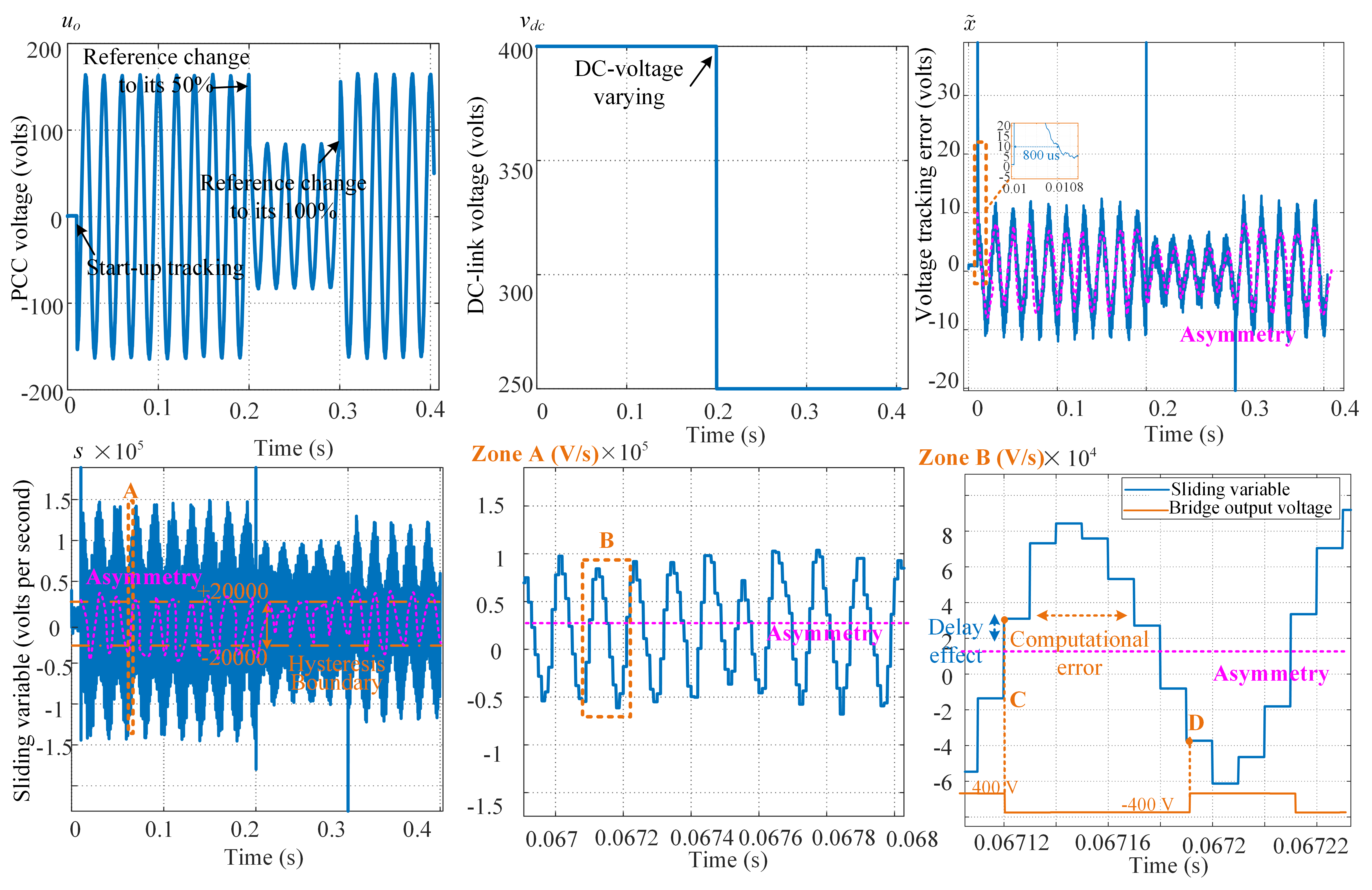}
	\caption{The simulation result of asymmetry when $t_{\text{di}} = 10\,\mu\mathrm{s}$, with a reference magnitude that steps from \(100\%\) to \(50\%\) at 0.2s with dc-voltage jump from 400V to 250V for examining  the $v_{dc}$ varying resilience and then back to \(100\%\) at 0.3s.  }
	\label{figs1}
	\end{figure*}
		
	\begin{figure*}[!t]\centering
		\includegraphics[width=0.88\textwidth]{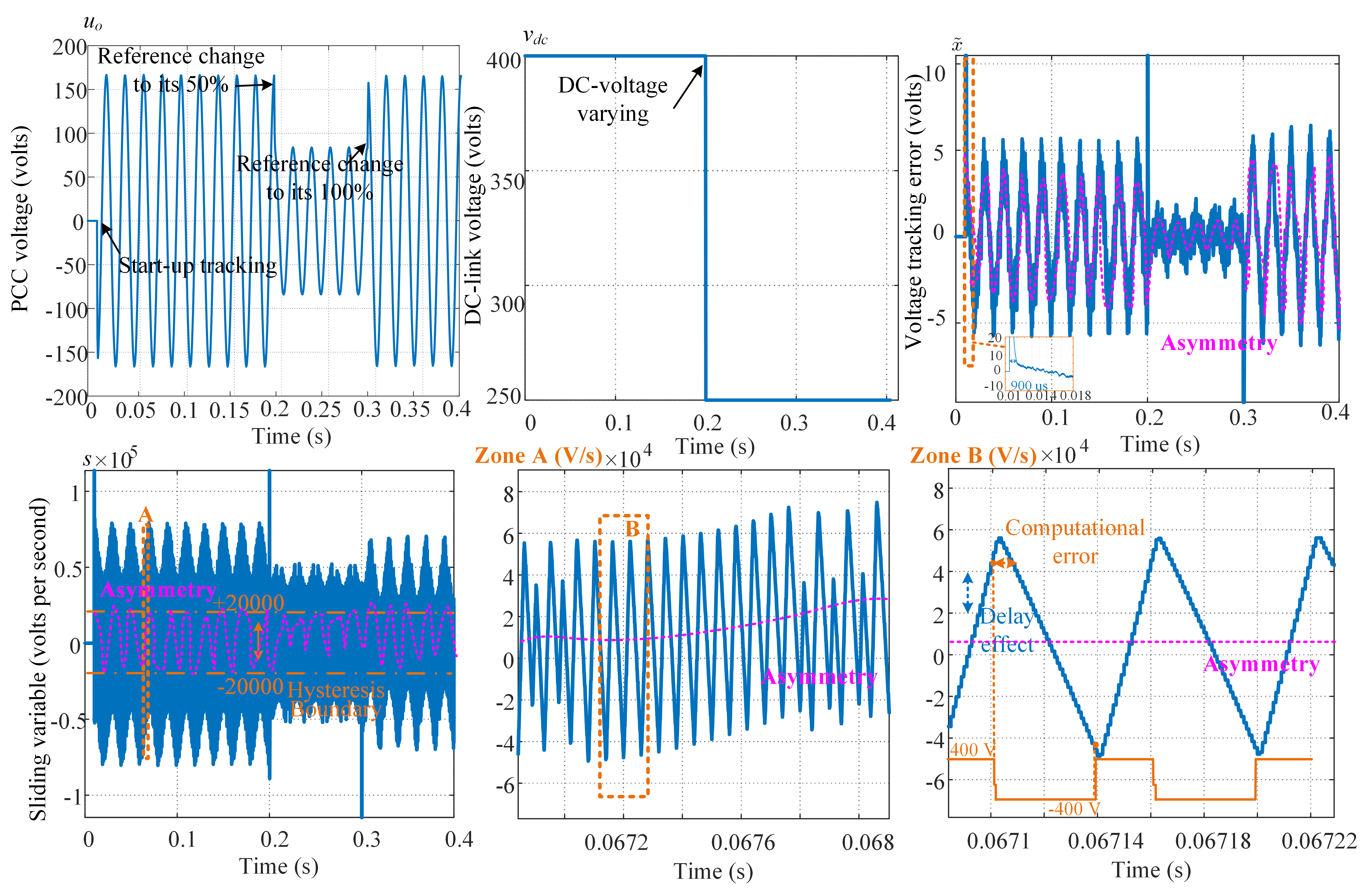}
		\caption{The simulation result of asymmetry when $t_{\text{di}} = 2\,\mu\mathrm{s}$, with a reference magnitude that steps from \(100\%\) to \(50\%\) at 0.2s with dc-voltage jump from 400V to 250V for examining $v_{dc}$ varying resilience and then back to \(100\%\) at 0.3s.}
		\label{figs2}
	\end{figure*}

	

Equation \eqref{eq9} and \eqref{eq10} show that the dynamic of $s$ is related to the decision interval $t_{\text {di}}$, namely the switching interval (combining the delay effect and computational error in Fig. \ref{figs1}), since the controller is a digital one. Now, let's consider the asymmetric dynamic of sliding mode control. The modeling is thought to be accurate, so $F$=0. Moreover, it is sufficient to require that $\eta$ is positive.
When the voltage state is close to sliding surface, the orders of $i_o$ magnitude is several tens of amperes, and the value of $L_f$ is several millihenries, the $u_o$ and  $i_o$ terms are the dominant term that affects the $s$ dynamics. 

Therefore, 
\begin{equation}
	\label{eqasmmetry}
	\dot{s}_{n-1} \;\approx\;
	-\frac{u_{o}}{L_f C_f}\;-\;\frac{1}{C_f}\,\frac{d i_o}{d t}\;\pm\;v_{dc}.
\end{equation}

When $\frac{x}{L_f C_f}+\frac{1}{C_f} \frac{d i_o}{d t}$ is at positive half cycle, the $s$ has an average down velocity tendency during a switching cycle because of the decision time or switching interval. Whenever the $\pm v_{dc}$ changes the sign, after a $t_{\text {di}}$ of the digital controller, the velocity of $s$ would change to another value. As a result, the $s$ dynamics is asymmetric across the $s$ surface.  

The simulation  verifies the claim above where a one phase inverter is controlled by a conventional sliding-mode and connected with a load. The parameter setting is on Table \ref{Simulationparameter}. 
Fig. \ref{controregion} shows the effectiveness of \eqref{eq:VGCR_final} to predict the voltage controllable region with $v_{dc}$ changing. By \eqref{eq:VGCR_final}, dc-link voltage level should be higher than $(110+20)*1.414= 181 volts$, which is coherent to the simulation result in Fig. \ref{controregion}, otherwise there is undesired loss of injected power and distorted output-voltage waveform. It also shows when the inverter is in the voltage precision region, the accuracy and the injected power to grid don't change anymore. This gives the explicit interaction mechanism between the synchronous reference signal and voltage formation loop with this boundary condition.
During $v_{dc}$ changes from 400V to 250V at 0.2 seconds, the condition \eqref{eq:VGCR_final} is always satisfied. In Fig.~\ref{figs1}, the $s$ isn't contained in the hysteresis boundary. Within the condition \eqref{eq:VGCR_final}, point C illustrates that it is not the system not precise, but it is that the $s$ plot is wrong, which is called the computational error. The delay effect arrow shows that the $t_{\text {di}}$ is also important to the error of  $s$. 

Contrary to the explanation in \cite{7} which attributes this effect to LC filter dynamics and in \cite{14} switching delay of transistors, the simulation in Fig.~\ref{figs1} shows that the delay effect ,  the computational error, and the potential asymmetric velocity of $s$ are reasons of the asymmetry of $s$ and asymmetric switching dynamics. 
Reducing  $t_{\text {di}}$ to about $2\,\mu\mathrm{s}$ in Fig.~\ref{figs2} lowers the computational error, consistent with the earlier qualitative prediction. The asymmetry caused by asymmetric switching dynamics creates a severe low-frequency-voltage error in  Fig.~\ref{figs1} and  Fig.~\ref{figs2}. 
If the voltage error at power-line frequency can be suppressed, with the 800 or 900 microseconds convergence speed, this control strategy would have a good voltage tracking performance for synchronization.

\begin{table}[!t]
	\renewcommand{\arraystretch}{1.3}
	\caption{Parameters of Simulation and Experimental Setups}
	\centering
	\label{Simulationparameter}
	\resizebox{\columnwidth}{!}{
		\begin{tabular}{c l c}
			\hline\hline \\[-3mm]
			Symbol & Description & Value  \\ \hline
			$f_n$  & Nominal frequency & $50$ Hz \\
			$S_n $ & Nominal power &  2 kVA  \\ 
			$V_n $ & Nominal one phase RMS voltage & 110 V \\
			$C_f$ & Filter capacitor & 330 $\mu$F\\ 
			$L_f$ & Filter inductor & 0.3 mH\\ 
			$v_{dc}$ & Nominal dc-link voltage & 290V\\
			$P_{sload}$ & Simulation active power  & 1600 W\\
			$Q_{sload}$ & Simulation reactive power  & 800 var\\
			$\lambda$ & Sliding control parameters & 4480 /s \\
			$h_{s}$           & Simulation hysteresis band            & 20000 V/s               \\
			$\omega_{0}$  & Natural frequency                   & 314.16 rad/s           \\
		
			$P_{\mathrm{step}}$ & Experiment active power change  & 1667.1 W               \\
			$Q_{\mathrm{step}}$ & Experiment reactive power change & 795.4 VAr              \\
		 
		$\zeta$       & Experiment damping ratio       & 2                  \\
			\hline\hline

		\end{tabular}	}
\end{table}

The reason why this asymmetry is important is that the smaller the $\Phi$ is, the smaller the $|\tilde{x}|$ is. This issue can be seen at these papers \cite{8391293}, \cite{8240975}, where there are low-frequency errors in output voltage whose rippling frequency is the grid frequency. Although the repetitive controller\cite{8240975} tries to change the sliding surface, but when an instant change occurs, it would lose fast convergence and accurate tracking, which would force the synchronous loop to interact more intensively with the inner voltage loop. Fortunately, according to the dominant terms in \eqref{eq9}, which is a low-frequency sinusoidal waveform, it is easier to correct the $s$ by a compensator extracting the asymmetry of $s$. 

\begin{figure}[!htbp]
	\includegraphics[width=3.2in]{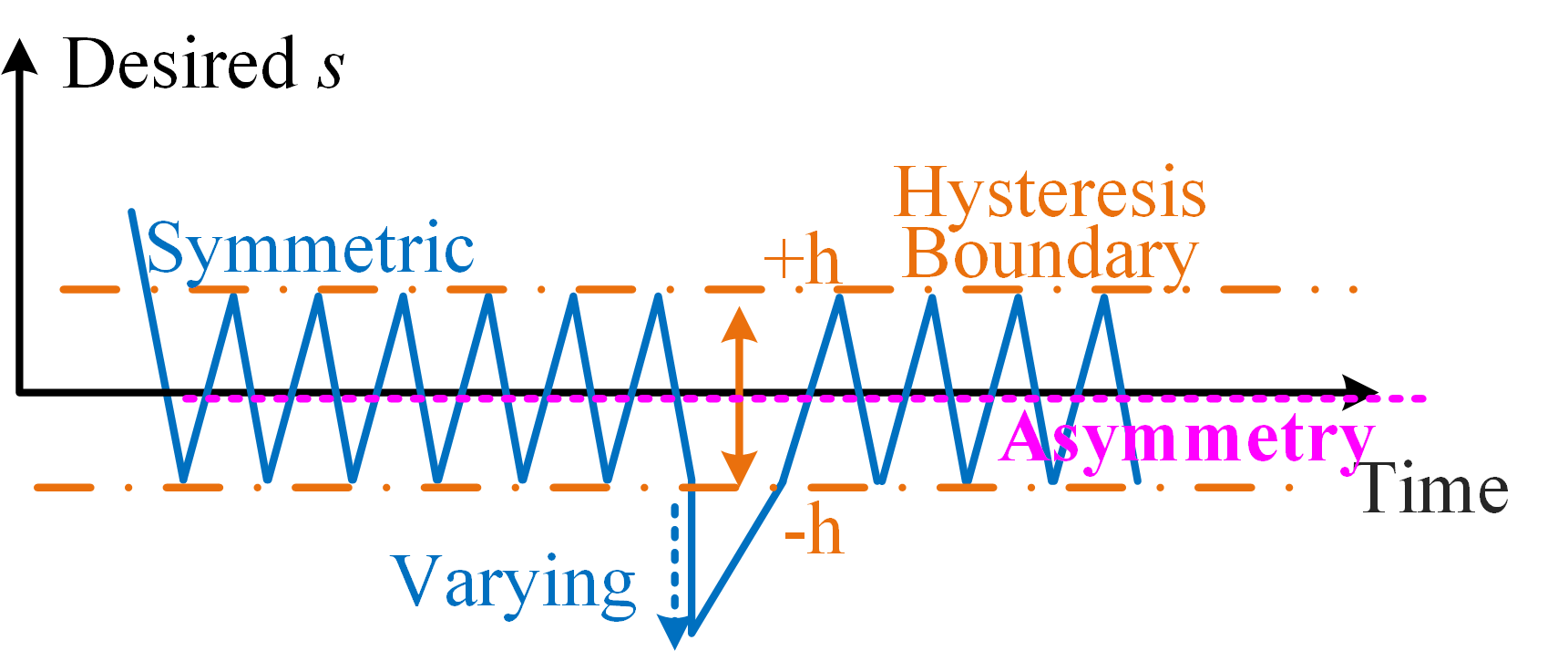}
	\centering
	\caption{The ideal $s$ dynamics, which seems a hysteresis dynamics with a bound.}
	\label{figds}
\end{figure}

With the symmetric notion above, the desired $s$ trajectory is given in Fig.~\ref{figds}.  The  $s$ dynamics is a function of sample time, control strategy, and physically allowed switching frequency. For example, the dead-band times, sampling process and derivative calculation can cause computation error and switching delay of power semiconductor devices \cite{14,15}. This causes the sliding variable to oscillate around the hysteresis band, creating a persistent offset from the ideal switching surface. Moreover, the asymmetry property of switching behavior in \eqref{eq9} can also damage the symmetry of $s$ variable.


 To address the  asymmetrical behavior, the sliding surface is decomposed in \eqref{eq3} into a symmetric dynamic component  $s_{\text{symmetric}}(t)$  which stands for the ideal symmetric sliding surface component shown in Fig.~\ref{figds}, and an error term $s_{\text{error}}$ which stands for the asymmetry. 
\begin{align}
\label{eq3}
\nonumber	s(x,t) &= s_{\text{symmetric}}  + s_{\text{error}} 
\\
&s_{\text{symmetric}}+s_{\text{error}} = \left( \frac{d}{dt} + \lambda \right) \tilde{x}
 \end{align}
 
 $x_{\text{comp}}(t)$ compensator is introduced to eliminate the asymmetry bias, thereby restoring the symmetry of the sliding variable $s$, and improving low-frequency tracking accuracy. This proof is shown in \eqref{eq4}
\begin{align}
	\label{eq4}
\nonumber	\tilde{x} &:= x - x_d + x_{\text{comp}} 
\\
\nonumber  & \text{s.t.} \quad \frac{d x_{\text{comp}}}{dt} + \lambda x_{\text{comp}} = s_{\text{error}}
 \\
&\text{Thus,} 	\quad \left( \frac{d}{dt} + \lambda \right)(x - x_d) = s_{\text{symmetric}}
\end{align}

To extract the low-frequency component of the sliding variable, a second-order band-pass filter is employed. The general form of the filter is given by:
\begin{equation}
	H(s) = \frac{2\zeta \omega_0 s}{s^2 + 2\zeta \omega_0 s + \omega_0^2}
	\label{eq:general_bpf}
\end{equation}
where $\omega_0$ is the natural (center) frequency in rad/s, and $\zeta$ is the damping ratio.

\begin{figure*}[!t]\centering
	\includegraphics[width=0.88\textwidth]{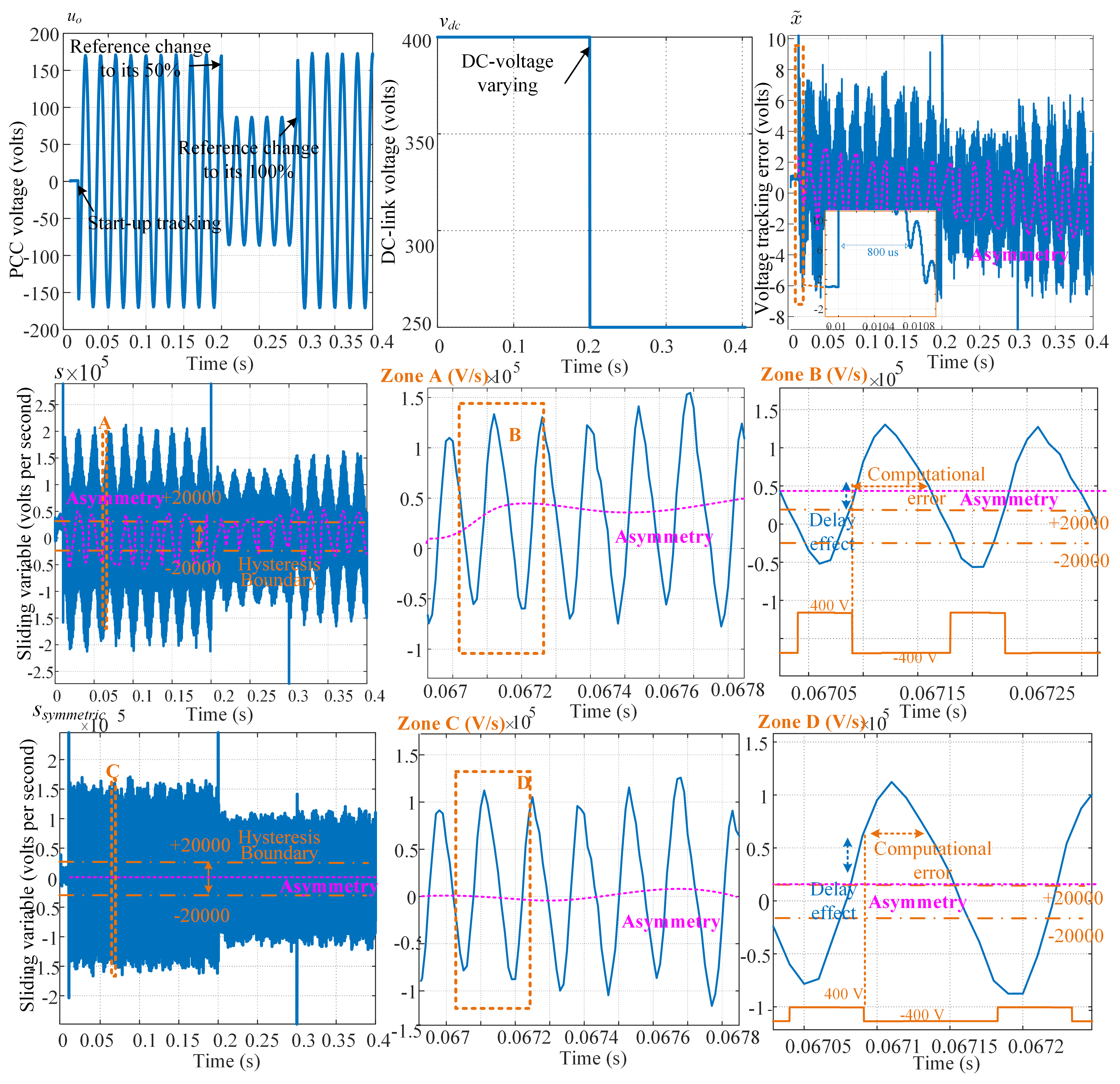}
	\caption{The simulation result of the proposed symmetric sliding-mode control when $t_{\text{di}} = 10\,\mu\mathrm{s}$, with a reference magnitude that steps from \(100\%\) to \(50\%\) at 0.2s with dc-voltage jump from 400V to 250V for examining  the $v_{dc}$ varying resilience  and then back to \(100\%\) at 0.3s.}
	\label{figs3}
	\end{figure*}
	\begin{figure*}[!t]\centering
		\includegraphics[width=0.88\textwidth]{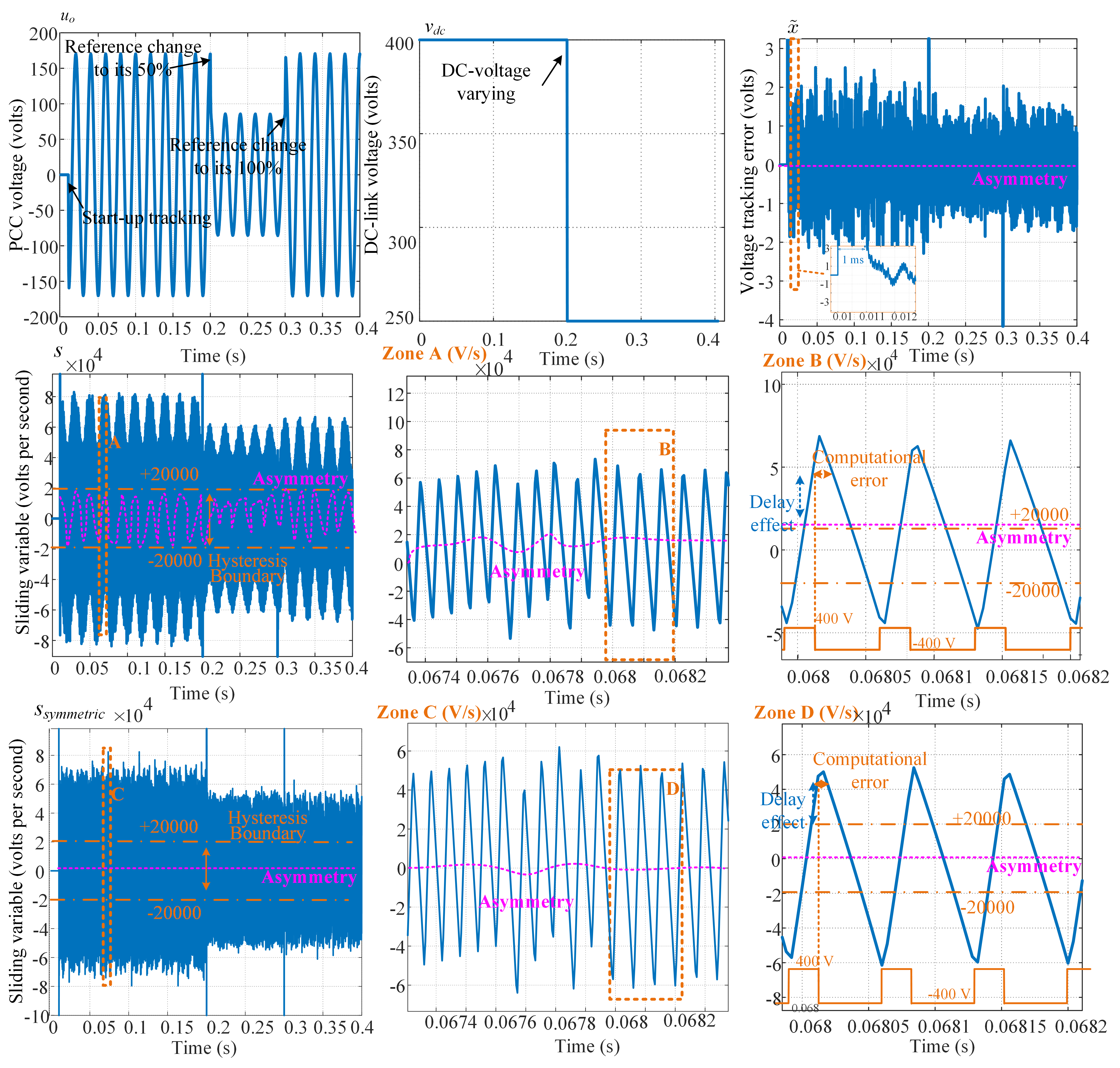}
		\caption{The simulation result of the proposed symmetric sliding-mode control when $t_{\text{di}} = 2\,\mu\mathrm{s}$, with a reference magnitude that steps from \(100\%\) to \(50\%\) at 0.2s with dc-voltage jump from 400V to 250V for examining the $v_{dc}$ varying resilience  and then back to \(100\%\) at 0.3s.}
		\label{figs4}
	\end{figure*}

This filter structure extracts the slow-varying envelope of the sliding variable, suppressing the low-frequency switching components. With the saturation input block at the input of this transfer function, this suppression function can only be activated when the sliding variable reaches the around the hysteresis band without losing the fast converging behavior outside the hysteresis boundary. Its effectiveness is shown in Fig. \ref{figs3} and  Fig. \ref{figs4}, which show a better tracking accuracy and the compensated $s$, namely $s_{\text{symmetric}}$ has a lower asymmetry.  Note that this compensator might cause a little damping-convergence effect if the $\zeta $ is small. So, the larger  $\zeta $ is better unless the  $s_{\text{fixed-error}} $ cannot be extracted efficiently.

 \section{Experimental Validation}
 
\begin{figure}[!htbp]
	\includegraphics[width=3.7in]{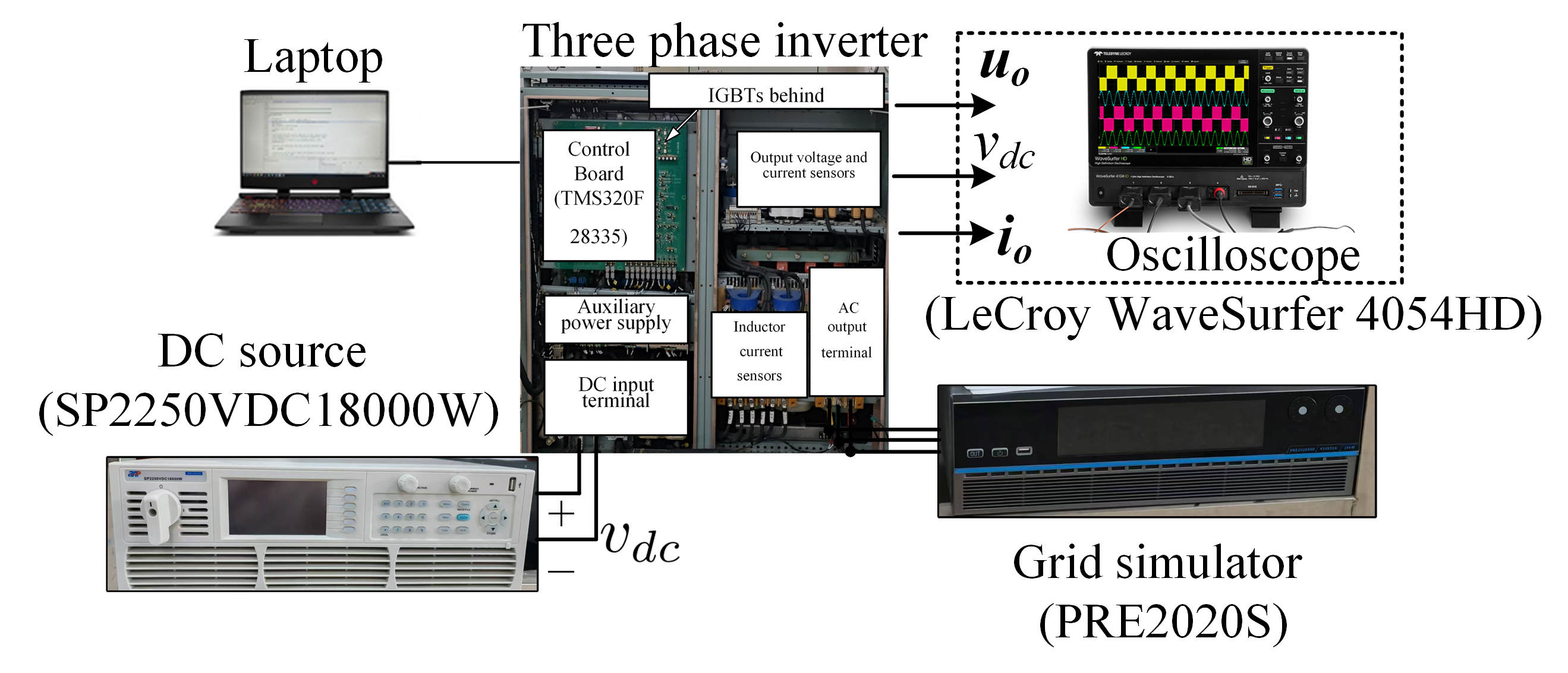}
	\centering
		\caption{Experimental setup.}
	\label{FIG_3}
\end{figure}

\begin{figure*}[!htbp]
 	\centering
 	\includegraphics[width=0.9\textwidth]{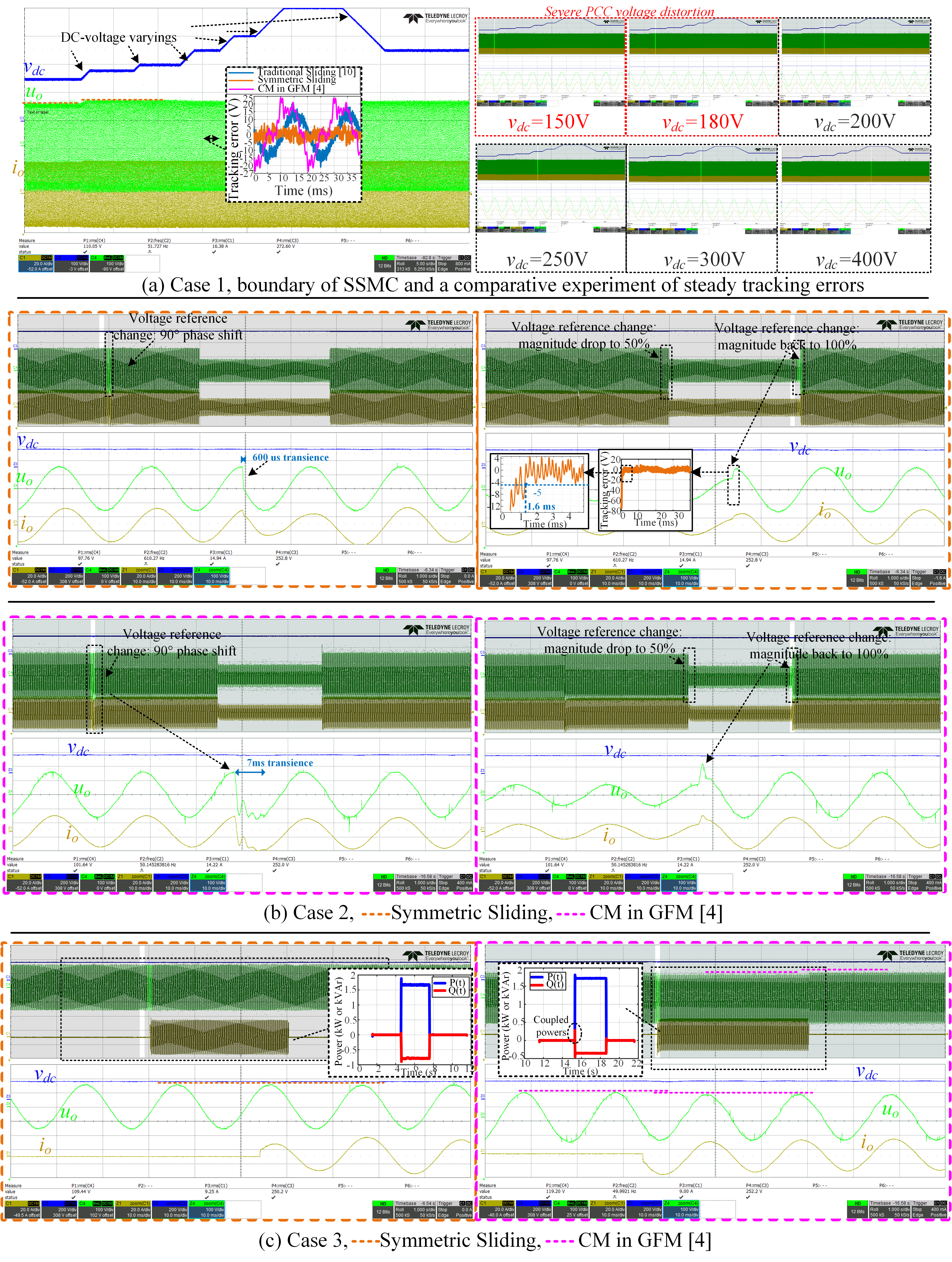}
 	\caption{Comparative experimental validation of the symmetric sliding-mode controller (SSMC).(a)~\textbf{Case 1-DC-link large varying:} The DC-link voltage is swept from 150V to 400V (150V and 180V are outside the precision region and then raised to 200V, 250V, 300V, and 400V). 
 		(b)~\textbf{Case 2-step reference voltage:} A sudden $90^{\circ}$ phase shift together with a 50\,\% magnitude drop and 100\,\% recovery. 
 		(c)~\textbf{Case 3-AC-side varying:} An abrupt 1.8\,kVA load change on the AC side.}
 	\label{experiment_time_domain}
 \end{figure*}

To verify the proposed symmetric sliding-mode control (SSMC) strategy, a series of experiments are conducted to evaluate the finite-time, almost no low-frequency error voltage tracking performance and its voltage precision region against ac-side and dc-link large varying. The experimental setup and procedures are designed to simultaneously demonstrate the effectiveness of tracking control and decoupling. The experiments are conducted on a laboratory-scale grid-forming inverter prototype  controlled by the proposed SSMC. Some experimental parameter setups are listed in Table \ref{Simulationparameter}, several of which remain unchanged from the simulation configuration. The inverter is connected to a programmable AC source, a controllable load bank, and a programmable DC supply to emulate AC power  and dc-link voltage variations, shown in Fig. \ref{FIG_3}. Three cases are examined and results are shown in Fig. \ref{experiment_time_domain}.

\begin{enumerate}[1)]
	
	\item \textbf{Case 1 DC-Link Large Varying Test}: A dc-link voltage changing from  150 volts to 180, both don't satisfy the condition \eqref{eq8} and have distortions of the output voltage. After increasing the voltage to 200, 300 and 400 volts, the output voltage is accurately controlled with a tracking error within $\pm$8\,V by SSMC, compared to methods in \cite{4,10752352}. The asymmetry suppressed compared with higher asymmetry of the traditional method without symmetry compensation in case 1, Fig. \ref{experiment_time_domain}. It shows a better decoupling effect of 38\% dc-link voltage changing of SSMC, compared to the compensated modulation (CM) method in \cite{4} and the effectiveness of the precision  region which can predict when the interaction intensifies.
	 
	\item \textbf{Case 2 Step Reference Voltage Test}:	Sudden changes (90 degrees phase change and voltage drop to its half and back to the nominal value) are applied to the voltage reference. This PCC voltage regulation task can be seen in many grid voltage drop conditions and in synchronization tasks of GFM \cite{1,2,3}. The response time could be within 600 microseconds with a smaller overshoot and damping than the CM method in \cite{4}. 

	\item \textbf{Case 3 AC-Side Varying Test} :A sudden current change (1.8kVA apparent power) is applied on the AC side. The voltage and current waveform are better in SSMC than direct modulation (DM) in \cite{4}. Also, the powers are obviously coupled by DM in \cite{4}. In contrast, the proposed SSMC achieves accurate tracking within $\pm$8\,V instantaneous error even under current variation. 
	
	The above proves a decoupled cross-impact between ac-side and dc-link varyings due to control decoupling and  superior performance of SSMC over conventional control under large operating condition varying. With the notion and explicit expression of voltage precision  boundary condition, the loss of voltage regulation can be predicted.

\end{enumerate}

\section{Conclusion}
This paper proposed a SSMC strategy that  can decouple ac-side power dynamics and dc-link voltage varyings from the voltage regulation task of GFMI. It also offers a nonlinear analysis based  control-parameter designing, which is physical and quantitative. After the investigation of the root causes of the asymmetry of sliding variable, by decomposing the sliding surface into a symmetric dynamic component and an error term, the proposed method effectively compensates for persistent tracking errors caused by the delay effect,  the computational error, and the potential asymmetric velocity of $s$ variable. Experimental results validate the rapid converging performance and dc-ac-varying resilience of the proposed method, particularly under changes in voltage reference, load, and dc-link voltage. In the explicit voltage precision  boundary, the proposed SSMC guarantees a fast, precise, and highly adaptive voltage tracking. This shrinks the coupling between the synchronous outer loop and the voltage inner loop, thereby increasing closed-loop robustness. In this region, the PCC voltage can be guaranteed as a controllable voltage source with a certain bounded error, so the inverters can be seen as a well-controlled black box. In this case, it helps model synchronization in large-scale integration scenarios involving numerous GFM inverters in future research.


\bibliographystyle{IEEEtranTIE}

\bibliography{BIB-TIE-0609} 

\begin{thebibliography}{10}
\providecommand{\url}[1]{#1}
\csname url@samestyle\endcsname
\providecommand{\newblock}{\relax}
\providecommand{\bibinfo}[2]{#2}
\providecommand{\BIBentrySTDinterwordspacing}{\spaceskip=0pt\relax}
\providecommand{\BIBentryALTinterwordstretchfactor}{4}
\providecommand{\BIBentryALTinterwordspacing}{\spaceskip=\fontdimen2\font plus
\BIBentryALTinterwordstretchfactor\fontdimen3\font minus
  \fontdimen4\font\relax}
\providecommand{\BIBforeignlanguage}[2]{{%
\expandafter\ifx\csname l@#1\endcsname\relax
\typeout{** WARNING: IEEEtran.bst: No hyphenation pattern has been}%
\typeout{** loaded for the language `#1'. Using the pattern for}%
\typeout{** the default language instead.}%
\else
\language=\csname l@#1\endcsname
\fi
#2}}
\providecommand{\BIBdecl}{\relax}
\BIBdecl

\bibitem{1}
D.~R{\'i}os-Castro, D.~P{\'e}rez-Est{\'e}vez, and J.~Doval-Gandoy, ``Ac-voltage
  controller for grid-forming converters,'' \emph{IEEE Transactions on Power
  Electronics}, vol.~38, no.~4, pp. 4529--4543, 2023.

\bibitem{2}
Z.~Li, C.~Zang, P.~Zeng, H.~Yu, S.~Li, and J.~Bian, ``Control of a grid-forming
  inverter based on sliding-mode and mixed ${H_2}/{H_\infty }$ control,''
  \emph{IEEE Transactions on Industrial Electronics}, vol.~64, no.~5, pp.
  3862--3872, 2017.

\bibitem{3}
D.~B. Rathnayake, S.~P. Me, R.~Razzaghi, and B.~Bahrani, ``H{$\infty$}-based
  control design for grid-forming inverters with enhanced damping and virtual
  inertia,'' \emph{IEEE Journal of Emerging and Selected Topics in Power
  Electronics}, vol.~11, no.~2, pp. 2311--2325, 2023.

\bibitem{4}
Z.~Zeng, P.~M. Gajare, D.~Divan, and M.~Saeedifard, ``Impact of dc voltage
  reference on subsynchronous dynamics in grid-forming inverters,'' \emph{IEEE
  Transactions on Power Electronics}, vol.~40, no.~7, pp. 8934--8938, 2025.

\bibitem{5}
J.~Liu, Y.~Xia, W.~Wei, Q.~Feng, and P.~Yang, ``Effect of control damping on
  small-signal stability of grid-forming vscs considering interaction between
  inner and outer loops,'' \emph{IEEE Transactions on Power Electronics},
  vol.~39, no.~6, pp. 7685--7695, 2024.

\bibitem{6}
T.~Liu and X.~Wang, ``Physical insight into hybrid-synchronization-controlled
  grid-forming inverters under large disturbances,'' \emph{IEEE Transactions on
  Power Electronics}, vol.~37, no.~10, pp. 11\,475--11\,480, 2022.

\bibitem{7}
L.~Zhao, Z.~Jin, and X.~Wang, ``Small-signal synchronization stability of
  grid-forming converters with regulated dc-link dynamics,'' \emph{IEEE
  Transactions on Industrial Electronics}, vol.~70, no.~12, pp.
  12\,399--12\,409, 2023.

\bibitem{8}
C.~Luo, X.~Ma, T.~Liu, and X.~Wang, ``Adaptive-output-voltage-regulation-based
  solution for the dc-link undervoltage of grid- forming inverters,''
  \emph{IEEE Transactions on Power Electronics}, vol.~38, no.~10, pp.
  12\,559--12\,569, 2023.

\bibitem{9}
A.~Tayyebi, A.~Anta, and F.~D{\"o}rfler, ``Grid-forming hybrid angle control
  and almost global stability of the dc-ac power converter,'' \emph{IEEE
  Transactions on Automatic Control}, vol.~68, no.~7, pp. 3842--3857, 2023.

\bibitem{10752352}
C.~Alfaro, R.~Guzman, A.~Camacho, {\'A}.~Borrell, and L.~G. de~{V}icu{\~n}a,
  ``A novel complex power sharing based on sliding mode control for islanded ac
  microgrids,'' \emph{IEEE Transactions on Industrial Electronics}, vol.~72,
  \href{http://dx.doi.org/10.1109/TIE.2024.3488330}{DOI
  10.1109/TIE.2024.3488330}, no.~6, pp. 5507--5517, 2025.

\bibitem{8391293}
J.~Fei, Y.~Zhu, and M.~Hua, ``Disturbance observer based fuzzy sliding mode
  control of pv grid connected inverter,'' in \emph{2018 5th International
  Conference on Electrical and Electronic Engineering (ICEEE)},
  \href{http://dx.doi.org/10.1109/ICEEE2.2018.8391293}{DOI
  10.1109/ICEEE2.2018.8391293}, pp. 18--22, 2018.

\bibitem{8240975}
L.~Zheng, F.~Jiang, J.~Song, Y.~Gao, and M.~Tian, ``A discrete-time repetitive
  sliding mode control for voltage source inverters,'' \emph{IEEE Journal of
  Emerging and Selected Topics in Power Electronics}, vol.~6,
  \href{http://dx.doi.org/10.1109/JESTPE.2017.2781701}{DOI
  10.1109/JESTPE.2017.2781701}, no.~3, pp. 1553--1566, 2018.

\bibitem{11}
Q.~Tang and L.~Peng, ``A slack bus grid-forming inverter based on symmetric
  sliding mode control against power sharing imbalances among
  microgenerators,'' in \emph{IECON 2024 - 50th Annual Conference of the IEEE
  Industrial Electronics Society}, pp. 1--6, 2024.

\bibitem{10}
B.~Lin, L.~Peng, K.~Yu, and H.~Xu, ``Harmonic disturbance suppression based on
  precise shaping of output impedance at selected frequencies for standalone
  voltage source inverter,'' \emph{IEEE Transactions on Industry Applications},
  vol.~59, no.~6, pp. 6963--6975, 2023.

\bibitem{Slotine1991AppliedNC}
\BIBentryALTinterwordspacing
J.-J.~E. Slotine and W.~Li, ``Applied nonlinear control,'' 1991. [Online].
  Available: \url{https://api.semanticscholar.org/CorpusID:106519536}
\BIBentrySTDinterwordspacing

\bibitem{10429007}
K.~Rayane, A.~Rabhi, B.~K. Oubbati, and M.~Benzoubir, ``Enhanced grid-forming
  inverter control through integral control in a cascaded framework,'' in
  \emph{2024 4th International Conference on Smart Grid and Renewable Energy
  (SGRE)}, \href{http://dx.doi.org/10.1109/SGRE59715.2024.10429007}{DOI
  10.1109/SGRE59715.2024.10429007}, pp. 1--6, 2024.

\bibitem{10329961}
P.~Alinaghi~Hosseinabadi, S.~Mekhilef, H.~R. Pota, and M.~Kermadi,
  ``Chattering-free fixed-time robust sliding mode controller for
  grid-connected inverters under parameter variations,'' \emph{IEEE Journal of
  Emerging and Selected Topics in Power Electronics}, vol.~12,
  \href{http://dx.doi.org/10.1109/JESTPE.2023.3336186}{DOI
  10.1109/JESTPE.2023.3336186}, no.~1, pp. 579--592, 2024.

\bibitem{464511}
S.~Saetieo, R.~Devaraj, and D.~Torrey, ``The design and implementation of a
  three-phase active power filter based on control,'' \emph{IEEE Transactions
  on Industry Applications}, vol.~31,
  \href{http://dx.doi.org/10.1109/28.464511}{DOI 10.1109/28.464511}, no.~5, pp.
  993--1000, 1995.

\bibitem{12}
\BIBentryALTinterwordspacing
Q.~Tang and L.~Peng, ``Theoretical grid-forming extreme of inverters,'' 2025.
  [Online]. Available: \url{https://arxiv.org/abs/2504.20367}
\BIBentrySTDinterwordspacing

\bibitem{14}
O.~Kukrer, H.~Komurcugil, and A.~Doganalp, ``A three-level hysteresis function
  approach to the sliding-mode control of single-phase ups inverters,''
  \emph{IEEE Transactions on Industrial Electronics}, vol.~56, no.~9, pp.
  3477--3486, 2009.

\bibitem{15}
\BIBentryALTinterwordspacing
S.~Sasitharan and M.~Mishra, ``Constant switching frequency band controller for
  dynamic voltage restorer,'' \emph{IET Power Electronics}, vol.~3, pp.
  657--667, 2010. [Online]. Available:
  \url{https://digital-library.theiet.org/doi/abs/10.1049/iet-pel.2008.0267}
\BIBentrySTDinterwordspacing

\end{thebibliography}

	
 \vspace{-1.4cm}
\begin{IEEEbiography}[{\includegraphics[width=1in,height=1.25in,clip,keepaspectratio]{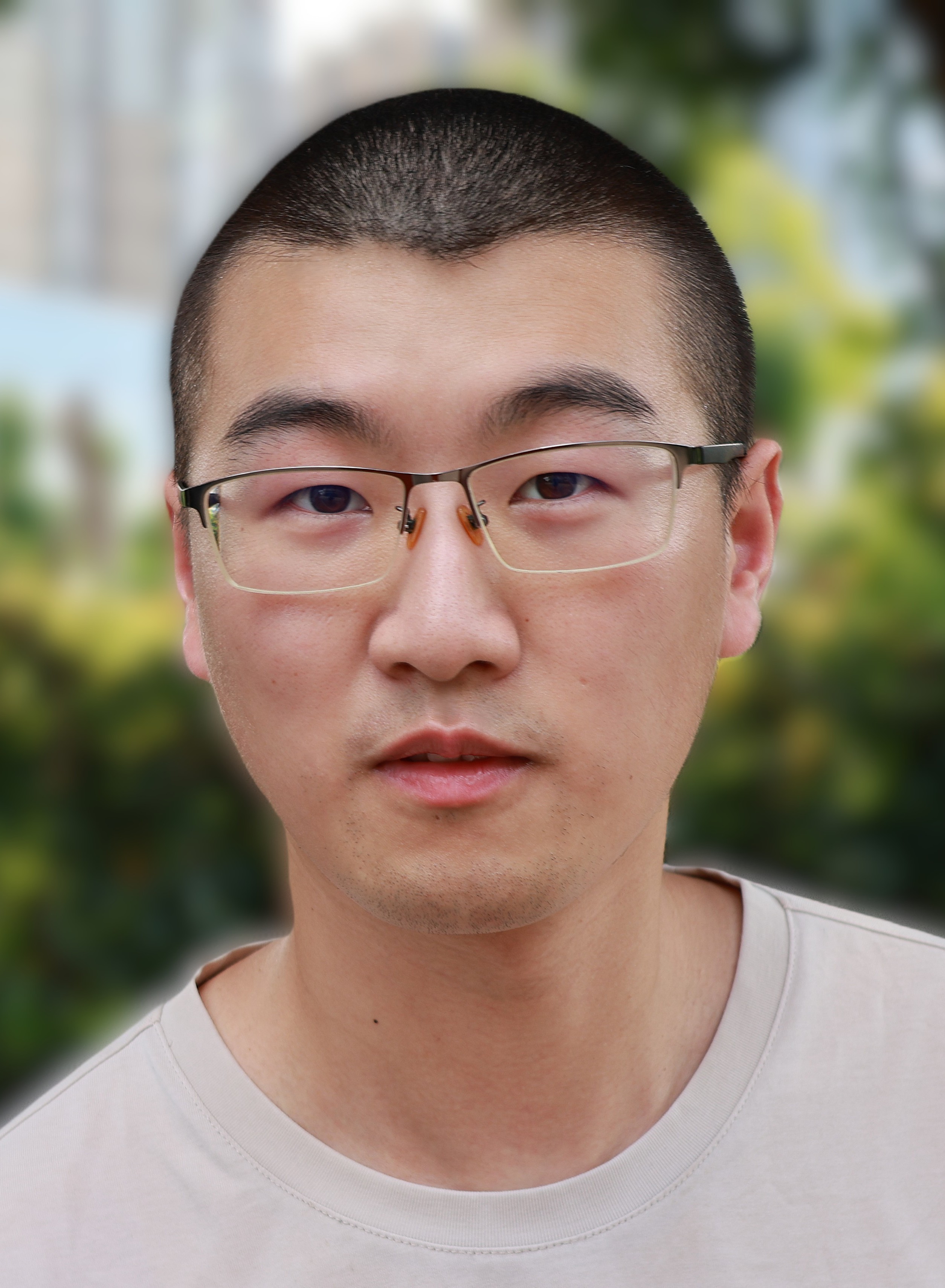}}]
	{Qianxi Tang} (Student Member, IEEE) received the B.Eng. degree from the
	School of Electrical Engineering, Beijing Jiaotong
	University (BJTU), Beijing, China, in July 2022.
	
	He is currently pursuing the Ph.D. degree with
	the State Key Laboratory of Advanced Electro
	magnetic and Technology, School of Electricaland
	Electronic Engineering, Huazhong University of
	Science and Technology (HUST), Wuhan, China.
	His research interests are controls in complex systems, networked systems with applications to smart power grids, the freedom and limitation of power electronics devices in power system.
  
  Dr. Tang is a reviewer for the IEEE Transactions on Industrial Electronics. He received the Best Presentation Recognition Award at IECON 2024, Chicago, IL, USA.

\end{IEEEbiography}

\vspace{-1.4cm}
\begin{IEEEbiography}[{\includegraphics[width=1in,height=1.25in,clip,keepaspectratio]{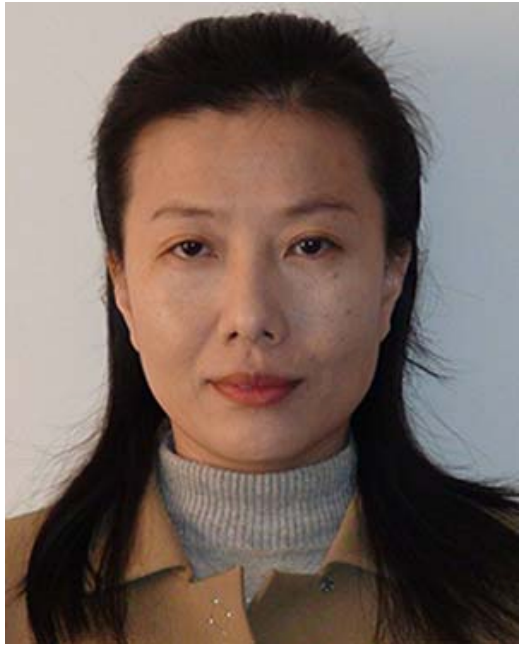}}]
	{Li Peng} (Senior Member, IEEE) received the B.S.,
	M.S., and Ph.D. degrees in power electronics
	from the Huazhong University of Science and
	Technology (HUST), Wuhan, China, in 1989,
	1992, and 2004, respectively.
	
	In 1992, she joined HUST, where she is
	currently a Full Professor of Power Electronics with the School of Electrical and Electronic Engineering. Her research interests include power electronic conversion, its control
	and applications, modular power supply and
	parallel control technique, renewable energy generation, power quality
	control, and modular multilevel converters for high-voltage direct-current
	applications.
\end{IEEEbiography}
\vspace{-1.4cm}
\begin{IEEEbiography}[{\includegraphics[width=1in,height=1.25in,clip,keepaspectratio]{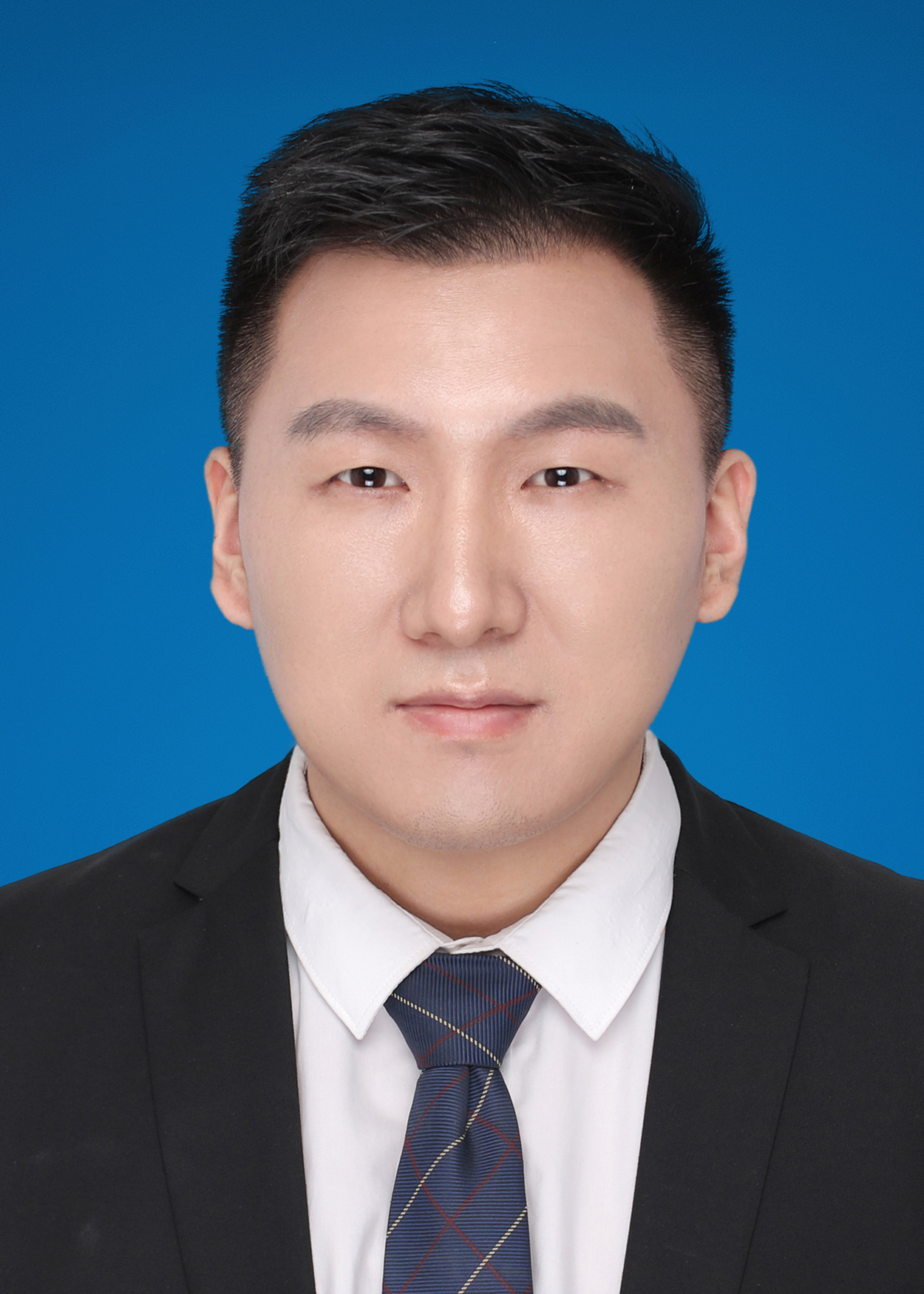}}]
	{Xuefeng Wang} received the B.Eng. degree in electrical engineering and automation from North China University of Water Resources and Electric Power, Zhengzhou, China, in 2013. 

He received the Ph.D. degree in electrical engineering with the School of Electrical and Electronic Engineering, Huazhong University of Science and Technology, Wuhan, China, in 2025. His major research interests include the control and analysis of power electronic converter.

\end{IEEEbiography}
\vspace{-1.4cm}
\begin{IEEEbiography}[{\includegraphics[width=1in,height=1.25in,clip,keepaspectratio]{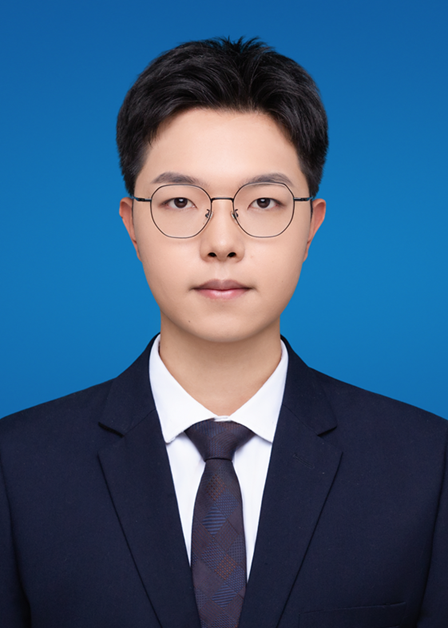}}]
	{Xinchen Yao} was born in Shaanxi, China, in 2001 and received the B.S. degree of electrical engineering from Huazhong University of Science and Technology, Wuhan, China, in 2023.
	 
	 He is currently working toward the M.S. degree with the School of Electrical and Electronic Engineering, Huazhong University of Science and Technology. His research interests include the grid-connected converters and stand-alone converters.
\end{IEEEbiography}

\end{document}